\PassOptionsToPackage{dvipsnames}{xcolor}

\documentclass[conference]{IEEEtran}

\usepackage{hyperref}
\usepackage[backend=bibtex,style=numeric,natbib=true,maxnames=40]{biblatex} %

\AtEveryBibitem{%
   \clearfield{pages}%
  \ifentrytype{misc}{%
  }{%
    \ifentrytype{unpublished}{%
    }{%
      \clearfield{url}%
      \clearfield{urldate}%
    }%
  }%
  \ifentrytype{book}{%
  }{%
   \clearfield{isbn}%
   \clearfield{issn}%
   \clearfield{series}%
  }%
}

\usepackage{adjustbox}
\usepackage{mathtools}
\usepackage{stmaryrd} %

\usepackage{physics}
\usepackage{amsmath}
\usepackage{tikz}
\usepackage{mathdots}
\usepackage{yhmath}
\usepackage{cancel}
\usepackage{color}
\usepackage{siunitx}
\usepackage{array}
\usepackage{multirow}
\usepackage{amssymb}
\usepackage{gensymb}
\usepackage{tabularx}
\usepackage{extarrows}
\usepackage{booktabs}
\usetikzlibrary{fadings}
\usetikzlibrary{patterns}
\usetikzlibrary{shadows.blur}
\usetikzlibrary{shapes}
\usepackage{pgf}
\usepackage{pgfplots}
\DeclareUnicodeCharacter{2212}{−}
\usepgfplotslibrary{groupplots,dateplot}
\usetikzlibrary{patterns,shapes.arrows}
\pgfplotsset{compat=newest}

\usepackage{subfigure}

\usepackage{dsfont}

\usepackage[frozencache,cachedir=.]{minted}

\usepackage{todonotes}

\usepackage{algorithm}
\usepackage[noend]{algorithmic}
\newcommand{\SWITCH}[1]{\STATE \textbf{switch} (#1)}
\newcommand{\ENDSWITCH}{\STATE \textbf{end switch}}
\newcommand{\CASE}[1]{\STATE \textbf{case} #1\textbf{:} \begin{ALC@g}}
\newcommand{\ENDCASE}{\end{ALC@g}}

\newcommand{\DEFAULT}{\STATE \textbf{default:} \begin{ALC@g}}
\newcommand{\ENDDEFAULT}{\end{ALC@g}}
\newcommand{\DEFAULTLINE}[1]{\STATE \textbf{default:} }
\usepackage{textcomp}
\usepackage{xcolor}
\def\BibTeX{{\rm B\kern-.05em{\sc i\kern-.025em b}\kern-.08em
    T\kern-.1667em\lower.7ex\hbox{E}\kern-.125emX}}
\begin{document}

\title{Living off the Analyst: Harvesting Features from Yara Rules for Malware Detection}

\author{\IEEEauthorblockN{Siddhant Gupta\IEEEauthorrefmark{1}, Fred Lu\IEEEauthorrefmark{1}\IEEEauthorrefmark{2}, Andrew Barlow\IEEEauthorrefmark{1}, Edward Raff\IEEEauthorrefmark{1}\IEEEauthorrefmark{2},\\ Francis Ferraro\IEEEauthorrefmark{1}, Cynthia Matuszek\IEEEauthorrefmark{1}, Charles Nicholas\IEEEauthorrefmark{1}, and James Holt\IEEEauthorrefmark{3}}
\\
\IEEEauthorblockA{\IEEEauthorrefmark{1}University of Maryland, Baltimore County\\
}
\IEEEauthorblockA{\IEEEauthorrefmark{2}Booz Allen Hamilton\\
}
\IEEEauthorblockA{\IEEEauthorrefmark{3}Laboratory of Physical Sciences\\
}
}

\maketitle
\begin{abstract}
  A strategy used by malicious actors is to ``live off the land,'' where benign systems and tools already available on a victim's systems are used and repurposed for the malicious actor's intent. In this work, we ask if there is a way for anti-virus developers to similarly re-purpose existing work to improve their malware detection capability. We show that this is plausible via YARA rules, which use human-written signatures to detect specific malware families, functionalities, or other markers of interest.
  By extracting \textit{sub-signatures} from publicly available YARA rules, we assembled a set of features that can more effectively discriminate malicious samples from benign ones. Our experiments demonstrate that these features add value beyond traditional features on the EMBER 2018 dataset. Manual analysis of the added sub-signatures shows a power-law behavior in a combination of features that are specific and unique, as well as features that occur often. %
  A prior expectation may be that the features would be limited in being overly specific to unique malware families. This behavior is observed, and is apparently useful in practice. In addition, we also find sub-signatures that are dual-purpose (e.g., detecting virtual machine environments) or broadly generic (e.g., DLL imports).  
\end{abstract}

\section{Introduction}

There often exists a gap between problems encountered in production, development, and deployment of anti-virus (AV) systems and those studied by academic researchers~\cite{BOTACIN2021102287,saul2024is,TirthCAMLIS}. This article is concerned with a problem that straddles this intersection: how do we get new static features for a malware detector? Static features are particularly necessary as they are one of the first tools used in production due to computational efficiency/throughput requirements, and thus, they need regular updates/evolution to stay effective as malware adapts and changes~\cite{Singh2012,203684,Joyce2021}. At the same time, human expertise and time to develop new features is expensive due to the high cost of analyst time~\cite{aonzo_humans_2023}. So we ask, \textit{is there any way to create new AV features more efficiently without overburdening our limited supply of malware analysts?} 
Inspired by the observation from ~\citet{10.1145/3291061} that analysts can often more quickly identify what makes a malicious sample malicious rather than determining a benign sample is benign, we look to re-purpose already existing activities as an additional ``boost'' to feature engineering. 

Specifically, analysts already develop signatures, relatively small and specific descriptors of a target family or functionality, as a regular part of their jobs. In this work, we hypothesize: \textit{Yara signatures~\cite{alvarez_victor_yara_2013} developed in the normal course of operations can be re-purposed to improve AV detectors for Windows PE files.} This re-purposing was non-obvious \textit{a priori}. Supporting this ``living off the land'' approach is that signatures are ubiquitous and incentive-aligned. 
Multiple works surveying analysts~\cite{Votipka2019}, reverse engineering AVs~\cite{Botacin2021a}, and working with professional analysts~\cite{Raff2020autoyara} have found that signature creation is still one of the key, common, and critical activities in real-world environments. This is often done when a specific malware family is not detected by current tools, which aligns well with our need to tackle concept drift and model degradation over time. However, all of these features are necessarily targeting malware, providing no new benign indicators, and are generally designed to be specific, meaning the approach is potentially incomplete in information. Even when code is reused between malware families~\cite{Calleja2019}, it does not mean that the content used in a signature will be reused. 

This article demonstrates that we can, in fact, extract useful features from Yara rules. Our work is organized as follows. First, work related to our own will be reviewed in \autoref{sec:related_work}. 
The approach we take to feature extraction and building our final model is detailed in \autoref{sec:yara_features}, which also analyzes why we take our approach. 
Our strategy at a high level is simple: we break Yara rules into their smaller components, which we term sub-signatures to avoid ambiguity, in order to increase their occurrence rate -- as a full Yara rule often fires too rarely and specifically to be useful as a feature. 
Then, we demonstrate two simple strategies for selecting these sub-signatures conditionally on the information already captured by an original model $f_0(\cdot)$ with its original feature set $\boldsymbol{z}$. 
The empirical validation of our method is shown in \autoref{sec:results}, ultimately demonstrating a relative improvement of 1.8\% at a low false-positive rate of 0.01\%.  This section includes a manual and statistical analysis of our choices and their impact on results, the features selected, and how the Yara features are meaningfully distinct from the existing EMBER features. Finally, we conclude in \autoref{sec:conclusion}.

\section{Related Work} \label{sec:related_work}

In a survey of 491 papers on malware research, \citet{BOTACIN2021102287} noted that little malware research deals with real-world observational challenges identified by users and companies.
There are issues with updating a model for real-world environments~\cite{Raff2020d,wu2024stabilizing}
\textit{A priori}.  We expect that most features extracted from Yara signatures will only be malicious indicators. This is not well aligned with standard machine learning, where features from both the benign and malicious classes are desired. However, recent works applied to malware detection have found forcing a model to use only positive indicators~\cite{Fleshman2018a,Incer:2018:ARM:3180445.3180449} can be a way of forcing robustness against a class of additive adversarial attacks,  the intuition being that only malware attempts to evade detection, and so features should only detect malware with a default answer of benign. Though this supports our approach in the abstract, other works have loosened the strict``positive features only'' requirement to improve accuracy and retain or expand the scope of robustness guarantees~\cite{NEURIPS2023_3ba82362,pfrommer2023asymmetric,NEURIPS2021_9fd98f85}. As the next paragraph will note, signatures are fallible, and so such techniques should be used to develop robust models, but the study of robustness is beyond the scope of this article. 

AutoYara~\cite{Raff2020autoyara}, YarGen \cite{Roth2013} and VxSig \cite{Blichmann2008} are among the many previous efforts to generate  Yara 
signatures~\cite{Kim2004,Newsome:2005:PAG:1058433.1059393,Griffin2009a,li_automatically_2024}. 
While we will show that features we predict to be useful can be harvested from Yara rules, it is important to note that Yara rules themselves can be attacked and subverted~\cite{Perdisci2006,Newsome:2006:PTS:2166373.2166380,li_packgenome_2023,coscia_yamme_2023}. As such, our method is not a panacea to adversarial attacks and defense, and indeed, such a discussion is larger than the scope of this article. 

Broadly, the complexity of malware detection and the impossibility of unbiased sampling of the population at large~\cite{agtr} have led to a wide variety of papers looking at alternative featurization techniques to meet different challenges. This includes custom dynamic analysis for less-common malware vectors like C\#~\cite{Botacin2017}, byte features, and miss-assumptions about packing~\cite{aghakhani_when_2020}, using the file path~\cite{Nguyen2019_filename_malicious,Li2017,Kyadige2019}, and using co-occurrence of other files~\cite{Ye:2011:CFC:2020408.2020448,Tamersoy:2014:GAL:2623330.2623342,Kwon:2015:DEI:2810103.2813724}, amongst many others. Our work continues this long-term trend of looking at alternative means of obtaining predictive information for malware detectors, targeting simplicity in fitting into existing processes. 

\section{Yara Rules as Features} \label{sec:yara_features}

\subsection{Extracting Yara Sub-signatures}
Yara rules are generally hand-written by analysts to detect or identify specific behaviors present in a class of malicious files.
Because of the flexibility of the Yara grammar
and the absence of a unifying intent (e.g., malware family version vs. functionality or capability detection) for using Yara Rules,
the rules can vary in robustness and looseness in identifying a given malware family \cite{canfora2020robustness}. Others (see for example \citet{10.1145/3460120.3484759}) have also observed a variety of intents and processes used by malware analysts.
Often, rules are designed to be specific,
targeting a particular malware family or interesting unit of functionality,
and so the desired fire rate of the rule is low with respect to the entire population of binary files.
In such cases,
using the entire Yara rule as a feature is problematic because a given rule may only fire on
few, if any, of the files in a corpus.

The crux of our approach will be to extract useful features from within given Yara rules.
It is necessary for us to have features that occur with some degree of regularity for them to be useful, and so we divide every Yara rule into \textit{sub-signatures: individual lines of the original signature that identify one specific component of the larger rule}, as illustrated in \autoref{fig:signature_example}. %
We discard any conditional logic that is used to combine the sub-signatures into the holistic rule as originally developed by an analyst. 

\begin{figure}[t]
    \centering
\begin{minted}[highlightlines={4-23},linenos=false,breaklines,breakanywhere,fontsize=\small]{cpp}
rule anubi
{
    strings:
        $av0 = "/c \"wmic product where name=\"ESET NOD32 Antivirus\" call uninstall /nointeractive \""
        $av1 = "/c \"wmic product where name=\"Kaspersky Anti-Virus\" call uninstall /nointeractive \""
        $av2 = "/c \"wmic product where name=\"Kaspersky Internet Security\" call uninstall /nointeractive \""
        $av3 = "/c \"wmic product where name=\"Avira Connect\" call uninstall /nointeractive \""
        $av4 = "SOFTWARE\\Microsoft\\Windows Defender\\Reporting"
        $cmd0 = "Invmod of %
        $cmd1 = "Expected : %
        $cmd2 = "Result :"
        $cmd3 = "%
        $anti0 = "DisableAntiSpyware"
        $anti1 = "DisableRoutinelyTakingAction"
        $anti2 = "TaskbarNoNotification"
        $anti3 = "DisableNotificationCenter"
        $ransom0 = "Dele"
        $ransom1 = "te S"
        $ransom2 = "hado"
        $ransom3 = "ws /"
        $ransom4 = "All"
        $ransom5 = "/Qui"
        $ransom6 = "et &"
    
    condition:
        (uint16(0) == 0x5A4D) and (2 of ($av*)) and (3 of ($cmd*)) and (3 of ($anti*)) and (6 of ($ransom*))
}
\end{minted}
    \caption{Example of a Yara signature from \url{ https://github.com/tjnel/yara_repo/blob/master/ransomware/crime_win_ransom_anubi.yara}. The highlighted ``strings'' of the Yara rule will be extracted and treated as 20 different features. The conditional logic that combines these components is ignored in our approach because it makes the whole rule as a singular unit too rare to be useful to a machine-learning model. By focusing on the individual components that we term sub-signatures, we obtain sufficient frequency in occurrence that they aid a predictive model.  }
    \label{fig:signature_example}
\end{figure}

We surveyed 22 GitHub repositories that hosted Yara 
rules for public consumption, listed in Appendix~\ref{sec:yara_repositories}. We found wide disparities in how rules are developed and used in each repository. 
For example, some analysts may use the Unix \texttt{strings} command as their main tool for writing signatures, while others hand-tune rules using tools built from YarGen or VxSig.
In addition, analysts may have preferences for different kinds of features or markers,
and even how their rules are organized within Yara (e.g., one ``uber signature'' that detects many different items of interest, or breaking each target of interest into its own Yara rule file). 
This was also documented by experiments in \cite{Raff2020autoyara},
in which three analysts each had different approaches for signature creation. 
Because our intent is to design a solution that is invariant to how analysts use our tools,
we did not place any restrictions on how or by whom the rules are written.
Nor did we attempt to filter the rules in each repository.
In total, we ended up with a set of 19,627 sub-signatures.

The set of sub-signatures were then turned into potential features by identifying whether each sub-signature occurred within each file,
over a corpus of benign and malicious files.
For this work, we used the 2018 version of the Ember dataset~\cite{anderson2018ember}.
The dataset is organized with a time-based split between train and test to avoid information leakage~\cite{235493},
with all malware collected in 2018.
There are 600k training and 200k testing files, evenly split between benign and malicious.
In other words,
for a given file $i$ and a set of $d$ sub-signatures,
we denote $\boldsymbol{x}_i \in \{0,1\}^d$ to be its corresponding feature vector.
In addition, we use $y_i \in \{0, 1\}$ to denote whether a given file is malicious,
with $y=1$ for malware.

\subsection{Identifying Useful Sub-signatures}

\begin{figure}[!h]
    \centering
    \subfigure[Distributions of accuracy, precision, and recall over Yara features. Short black lines indicate the extrema of each distribution. Long black lines indicate the median.]{
        \input{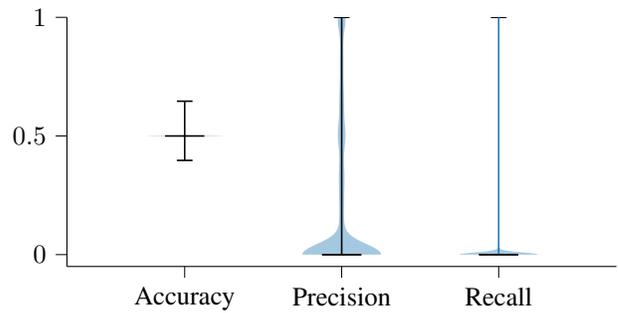}
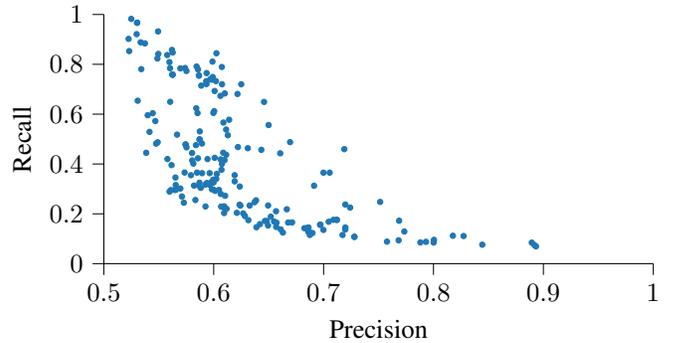 \label{fig:yara_stats}
    }
    \subfigure[Precision vs. recall of the top 200 Yara features selected by individual accuracy]{
        \begin{tikzpicture}

\definecolor{darkgray176}{RGB}{176,176,176}
\definecolor{steelblue31119180}{RGB}{31,119,180}

\begin{axis}[
width=0.49\textwidth,
height=0.27\textwidth,
axis x line* = bottom,
axis y line* = left,
tick align=outside,
tick pos=left,
x grid style={darkgray176},
xlabel={Precision},
xmin=0.5, xmax=1,
xtick style={color=black},
y grid style={darkgray176},
ylabel={Recall},
ymin=0, ymax=1,
ytick style={color=black}
]
\addplot [draw=steelblue31119180, fill=steelblue31119180, mark=*, mark size=1, only marks]
table{%
x  y
0.64581391723685 0.649066666666667
0.625041237650409 0.71996
0.602462194187966 0.844326666666667
0.71886130275392 0.459503333333333
0.607381775648195 0.789296666666667
0.598879278720329 0.810816666666667
0.621619976936582 0.681003333333333
0.649947818501846 0.556346666666667
0.607568908629085 0.719983333333333
0.602017562600317 0.732643333333333
0.599074372762246 0.749036666666667
0.669490323141627 0.487636666666667
0.610059129316351 0.68301
0.598195911091906 0.738313333333333
0.593539958088635 0.764736666666667
0.596800593438314 0.74017
0.60611552344404 0.67336
0.600827196318947 0.69293
0.593332072003892 0.731736666666667
0.593332072003892 0.731736666666667
0.584438419314564 0.79109
0.585751727663157 0.779233333333333
0.593395221678122 0.720603333333333
0.586569298441038 0.75477
0.58656625944232 0.75477
0.588662746788895 0.714493333333333
0.660512366612632 0.442776666666667
0.61404049638906 0.577603333333333
0.705525819146812 0.365116666666667
0.699888847721378 0.365206666666667
0.600436245671063 0.612026666666667
0.643339279008637 0.456856666666667
0.608820145999004 0.56629
0.573904484869747 0.785233333333333
0.575204234693245 0.773333333333333
0.599697226115239 0.604766666666667
0.611320597566424 0.538243333333333
0.631008371315985 0.46332
0.569659547760004 0.78358
0.612843891223341 0.51532
0.562007835327105 0.85786
0.562694589445314 0.84763
0.622047139419775 0.467926666666667
0.584123078460333 0.623553333333333
0.585291829045038 0.604253333333333
0.557719165414018 0.837283333333333
0.691331098257378 0.312746666666667
0.559735891366638 0.80874
0.560283012352618 0.78423
0.562488261982662 0.758736666666667
0.562488261982662 0.758736666666667
0.562485481997677 0.758736666666667
0.562484092015489 0.758736666666667
0.549365735573825 0.931973333333333
0.751478871103111 0.24772
0.611355483928147 0.436876666666667
0.608656208828531 0.445043333333333
0.587307676701353 0.530716666666667
0.54959364811276 0.84171
0.609825189708062 0.41548
0.587211150219544 0.49883
0.587058777337155 0.498393333333333
0.606319562528984 0.418376666666667
0.548789076472248 0.823386666666667
0.589346827026498 0.48218
0.71977955303873 0.237263333333333
0.600913797493422 0.42481
0.60748954165381 0.400313333333333
0.560423851421898 0.649473333333333
0.724149028610661 0.22501
0.583865317881228 0.475593333333333
0.619027690590779 0.355
0.594354720686611 0.41966
0.607092990784697 0.376826666666667
0.619088682263547 0.330246666666667
0.574378032827545 0.478596666666667
0.57446953482593 0.477763333333333
0.585662576063972 0.422836666666667
0.537403800561695 0.88404
0.580297295767041 0.444126666666667
0.602816539918651 0.360136666666667
0.623681700321094 0.309483333333333
0.575249599408357 0.4667
0.566661195309223 0.517843333333333
0.768705458765417 0.172223333333333
0.596317420004598 0.363153333333333
0.602329136637567 0.340153333333333
0.580443099171817 0.41421
0.581479176015866 0.40168
0.599619492419483 0.33723
0.533505650011612 0.88823
0.530308072035835 0.96684
0.530306243715148 0.966836666666667
0.638528084147761 0.254153333333333
0.589436261512321 0.363096666666667
0.666530986058818 0.218333333333333
0.599833043819859 0.325743333333333
0.598268561097832 0.328716666666667
0.649522961809129 0.23351
0.637178736305381 0.246596666666667
0.584827740754771 0.365363333333333
0.712024205094956 0.175713333333333
0.712031771332856 0.175703333333333
0.529890189129376 0.921023333333333
0.708962143711448 0.175853333333333
0.604791888134762 0.29584
0.596448774355751 0.316316666666667
0.593822993481797 0.323716666666667
0.656715020083665 0.210363333333333
0.534112446886397 0.7806
0.592419806913734 0.317043333333333
0.544545227694315 0.60375
0.599753553608824 0.2969
0.610398440009862 0.272333333333333
0.601333461374892 0.29222
0.546901177864555 0.572346666666667
0.598292082017453 0.298693333333333
0.606467150750716 0.279516666666667
0.632741563897547 0.23345
0.704631505413902 0.168113333333333
0.579315859907954 0.35497
0.58703819661866 0.325
0.623549257166955 0.236576666666667
0.57351595786392 0.365683333333333
0.524972476475849 0.98229
0.62427978024923 0.23295
0.586641540515918 0.31338
0.588190549861407 0.304153333333333
0.773451716759654 0.128926666666667
0.719774403757732 0.145063333333333
0.697142941997565 0.15649
0.58210582390763 0.313246666666667
0.539923839458414 0.5955
0.69676216054909 0.155656666666667
0.6519658710942 0.188483333333333
0.827339203104014 0.11088
0.557912269976699 0.419806666666667
0.549015630458546 0.48718
0.817626293365794 0.111946666666667
0.561656221006192 0.395503333333333
0.671318647139167 0.165473333333333
0.547568333762867 0.48159
0.719754816112084 0.136993333333333
0.667710625430295 0.164873333333333
0.609472364913767 0.230173333333333
0.626842465327446 0.20158
0.541596261439783 0.528863333333333
0.609299010186864 0.225093333333333
0.606754178297587 0.228956666666667
0.611667114143668 0.220226666666667
0.654949312326467 0.169053333333333
0.56518090214009 0.345696666666667
0.621038531402535 0.203996666666667
0.686277600037824 0.14515
0.657143617871026 0.16454
0.648244892796155 0.170823333333333
0.628421767812918 0.191153333333333
0.522568710257346 0.902063333333333
0.646655425810348 0.171303333333333
0.699844800827729 0.13528
0.684383412383028 0.14237
0.682539682539683 0.14233
0.530837737564643 0.65387
0.523116424354425 0.852986666666667
0.889436099482886 0.08428
0.56967189890633 0.30159
0.565435359353713 0.31543
0.583337775577809 0.255336666666667
0.609474737368684 0.203056666666667
0.631554188529039 0.17453
0.800463583556747 0.0955433333333333
0.592526629153546 0.229556666666667
0.649435691454756 0.153063333333333
0.656861143697192 0.14745
0.642064080944351 0.158643333333333
0.717247115084728 0.114986666666667
0.686072099472225 0.127826666666667
0.565568972757002 0.296316666666667
0.68776511379145 0.124306666666667
0.689884192999196 0.122916666666667
0.564508020948005 0.29571
0.7280254489667 0.107563333333333
0.728009024252679 0.107563333333333
0.57110869380951 0.26951
0.660941730618261 0.137513333333333
0.891852029020171 0.0745766666666667
0.768259348349005 0.0936866666666667
0.793118032687834 0.08751
0.56110896408411 0.29522
0.538574823577829 0.44494
0.800094354458248 0.0847966666666667
0.638871893482504 0.146186666666667
0.687451503153539 0.115173333333333
0.788051867860451 0.0850833333333333
0.572813412216481 0.244633333333333
0.84438690586277 0.0760933333333333
0.559811390127009 0.28929
0.662936864841897 0.124813333333333
0.893249742356579 0.06934
0.757673621034581 0.08837
};
\end{axis}

\end{tikzpicture} \label{fig:yara_pvr}
    }
    \caption{Distributional statistics of Yara sub-signatures over the Ember dataset in terms of predicting whether a file is malware. This shows that individual sub-signatures have little predictive power and tend to be highly specific in what they detect. Thus, we need to consider a joint feature selection to obtain meaningful results. }
\end{figure}

Due to the large number of Yara sub-signatures which we were able to collect,
it is unlikely that all of them will make useful features.
Our next challenge is to identify useful Yara features for malware discrimination.
It is infeasible to manually inspect such a large number of sub-signatures to pick out relevant features.
Nor is it ideal to keep all of the features in a model if most of them turn out to be ineffective or redundant.

To emphasize the difficulty, we compute the discriminative performance of each sub-feature (19,627 extracted from the 22 repositories) over the Ember training set in terms of three metrics: accuracy, precision, and recall,
as shown in Fig.~\ref{fig:yara_stats}.
We note that the accuracies of all the sub-signatures fall within a narrow range between 40\% and 65\%,
while the vast majority are only slightly above 50\%. 
Although precision and recall metrics have a wider range, ranging from 0 to 1,
the majority of precision  and recall scores are only slightly above 0. 
As a result, we suspect that many of the features may detect a certain rare behavior with high precision but fail to detect most malware files,
or conversely,
flag a large proportion of files as malicious with a high false positive rate~\cite{Nguyen2021,Nguyen2022}.
Furthermore, the existence of features with sub-50\% accuracy raises the concern that certain sub-signatures may have low specificity,
being triggered by benignware more often than the actual malware files being targeted.

To attempt to filter out such behaviors, one could consider selecting the highest-accuracy rules.
For example, the top 200 rules by accuracy fall between 53\% and 65\%.
Even among such rules, however,
there is a wide range of precision and recall, as shown in Fig.~\ref{fig:yara_pvr}.
This indicates that such rules may still return only one class much of the time.
Thus filtering sub-signatures by their individual metrics on the Ember dataset may not effectively select useful features.

In spite of the above challenges, we will show that Yara sub-signatures are in fact useful features for malware classification,
when used in combination with principled feature selection.
In our method, we propose using $L_1$-regularized (also known as Lasso) logistic regression to perform automatic feature selection.
This model minimizes the objective
\begin{equation} \label{eq:lasso}
    L(w) \coloneqq \frac{1}{n} \sum_{i=1}^n \ell(w, \boldsymbol{x}_i, y_i) + \lambda \lVert w \rVert_1
\end{equation}
This model can learn a linear classification model with weights $w$ while also keeping $w$ sparse.
By varying $\lambda$ we can adjust the number of non-zero components in $w$,
corresponding to the selected Yara features, and can be scaled to large corpora~\cite{10.1145/3637528.3672038,lu2024optimizingoptimalweightedaverage}.

In Section~\ref{sec:results}, we will show that (1) combining Yara sub-signatures as features can give high malware discrimination performance; (2) the Lasso-based selection outperforms selecting sub-signatures based on individual accuracy, precision, or recall.

\subsection{Model Learning on Yara Sub-signatures}

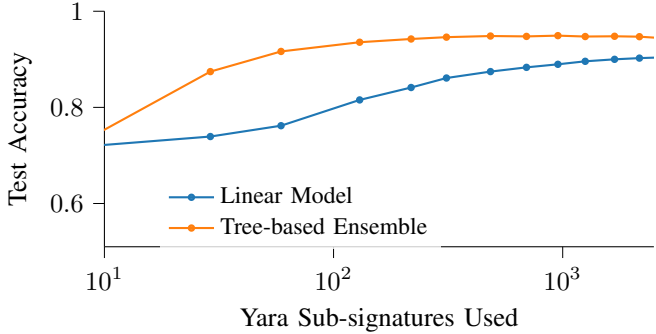
\begin{figure}[h]
    \centering
    \begin{tikzpicture}

\definecolor{darkgray176}{RGB}{176,176,176}
\definecolor{darkorange25512714}{RGB}{255,127,14}
\definecolor{lightgray204}{RGB}{204,204,204}
\definecolor{steelblue31119180}{RGB}{31,119,180}

\begin{axis}[
width=0.49\textwidth,
height=0.26\textwidth,
axis x line* = bottom,
axis y line* = left,
legend cell align={left},
legend style={
  fill opacity=0.8,
  draw opacity=1,
  text opacity=1,
  at={(0.1,0.3)},
  anchor=north west,
  draw=none
},
log basis x={10},
tick align=outside,
tick pos=left,
x grid style={darkgray176},
xlabel={Yara Sub-signatures Used},
xmin=10, xmax=2500,
xmode=log,
xtick style={color=black},
xtick={0.01,0.1,1,10,100,1000,10000,100000},
xticklabels={
  \(\displaystyle {10^{-2}}\),
  \(\displaystyle {10^{-1}}\),
  \(\displaystyle {10^{0}}\),
  \(\displaystyle {10^{1}}\),
  \(\displaystyle {10^{2}}\),
  \(\displaystyle {10^{3}}\),
  \(\displaystyle {10^{4}}\),
  \(\displaystyle {10^{5}}\)
},
y grid style={darkgray176},
ylabel={Test Accuracy},
ymin=0.51, ymax=1,
ytick style={color=black}
]
\addplot [thick, steelblue31119180, mark=*, mark size=1, mark options={solid}]
table {%
1 0.5
1 0.5
7 0.71602
29 0.73931
59 0.761855
130 0.815495
218 0.84132
311 0.86107
484 0.874435
695 0.88326
953 0.8896
1254 0.89583
1684 0.899975
2162 0.902475
2727 0.90404
3436 0.90653
4142 0.906825
4901 0.906765
5512 0.90633
5870 0.90636
};
\addlegendentry{\small Linear Model}
\addplot [thick, darkorange25512714, mark=*, mark size=1, mark options={solid}]
table {%
7 0.71293
29 0.874405
59 0.916365
130 0.93554
218 0.94228
311 0.946085
484 0.948445
695 0.947655
953 0.949245
1254 0.94738
1684 0.94792
2162 0.94707
2727 0.944005
3436 0.94712
4142 0.947605
4901 0.944595
5512 0.947415
5870 0.9469
};
\addlegendentry{\small Tree-based Ensemble}
\end{axis}

\end{tikzpicture}
    \caption{Using only Yara sub-signatures as features, we see that only a limited number of the original 19k sub-signatures are needed for predictive accuracy. This also shows that training a tree ensemble after feature selection provides higher accuracy, suggesting a two-step model-building process. This informs our strategy for the full approach of leveraging Yara sub-signatures.}
    \label{fig:lr_vs_tree}
\end{figure}

While a regularized linear classifier can efficiently perform feature selection,
we find that prediction accuracy can be further improved by switching to a more complex tree-based ensemble.
In Fig.~\ref{fig:lr_vs_tree},
we take the features selected by the Lasso model (Eq.~\ref{eq:lasso}) for each value of $\lambda$ and then fit an XGBoost model over those features.
We see that for any given set of Yara features,
the tree-based model achieves higher accuracy than the linear model. 

This higher accuracy indicates that there are non-linear interactions among Yara features beyond simple additive effects.
This makes sense since we had previously decomposed Yara rules into sub-signatures,
disregarding the conditional logic that combined the sub-signatures into a single rule.
The tree-based model thus has the ability to ``reconstruct'' useful rules from the sub-signatures using the Ember training data. Note that we are not attempting to validate the ``correctness'' of the tree's induced rules, as the task has changed from some specific identification of the original signature to a more general ``benign vs malicious'' task. 

We also highlight that with just 300 Yara sub-signatures, the accuracy has hit a plateau.
This reinforces the notion that it will be useful to perform feature selection to identify the relatively small set of effective Yara sub-signatures for malware discrimination.

\subsection{Incorporating Side Information} \label{subsec:side_info}

Our method can incorporate prior existing features in addition to Yara sub-signatures. 
For example, the Ember 2018 dataset was published with a set of 2381 features extracted using the LIEF library\footnote{\url{https://lief.re/}} representing a reasonable set of engineered features\footnote{The set of features is intentionally restricted compared to the original company's commercial offering. This is a balance that research in this space must play to obtain real-world impact but not jeopardize the company's needs to maintain some competitive edge.}. 
We denote $\boldsymbol{z}$ as the existing side features, which is stacked as a matrix $Z$.

We also introduce 
\begin{enumerate}
    \item $f_0(\cdot)$, the baseline classification model trained on the original feature set $\boldsymbol{z}$, that is not derived from Yara rules (i.e., the Ember features in our testing). 
    \item $g(\cdot)$, the feature selection model, which returns the subset of Yara sub-signatures that are predicted to be useful (i.e., \autoref{eq:lasso} in our experiments).
    \item  $f(\cdot)$, the final malware detection model trained from original features $\boldsymbol{z}$ and the added sub-signatures from $g(\cdot)$. 
\end{enumerate}

The simplest approach to feature selection %
is \textit{independent selection}, which uses $g$ to select Yara features without any side information by solving Eq.~\ref{eq:lasso} directly (i.e, the same way of selecting features in \autoref{fig:lr_vs_tree}).
Then the selected Yara features from $g(\boldsymbol{x})$ are concatenated to the side features $\boldsymbol{z}$ to use as predictors for $f$.

We also consider two more tailored approaches.
First is \textit{conditional selection},
where we first use $f_0(\cdot)$ to obtain the predicted probabilities based on the original features. Then $f_0(\boldsymbol{z})$ is added as a single uni-variate feature to $\boldsymbol{x}$ to train $g$ by minimizing
\begin{equation}
    L_{CS}(w) = \frac{1}{n} \sum_{i=1}^n \ell(w, [f_0(\boldsymbol{z}_i), \boldsymbol{x}_i], y_i) + \lambda \lVert w \rVert_1
\end{equation}
This approach provides conditional information about the existing set of features $\boldsymbol{z}$ at minimal compute cost,
and uses predictions that are likely already available.

The second alternative is \textit{stacked feature selection}, where all sub-signatures and the original $\boldsymbol{z}$ are concatenated into one larger feature vector as input to $g$, which has higher computational cost but may allow more fidelity in feature selection. In other words, the selection model $g$ minimizes
\begin{equation}
    L_{SFS}(w) = \frac{1}{n} \sum_{i=1}^n \ell(w, [\boldsymbol{z}_i, \boldsymbol{x}_i], y_i) + \lambda \lVert w \rVert_1
\end{equation}

This gives us two sets of options to evaluate the effectiveness of feature selection. A priori, we expect that conditional selection will be easier to leverage in a commercial environment where larger (private) datasets may be used, and the computational cost of regular model updates is a non-trivial financial consideration. Testing the stacked feature selection helps us to determine if there may be predictive efficacy lost in using the prior approach. 

\subsection{Full Algorithm}

\begin{algorithm}[!h]
\begin{algorithmic}[1] %
\STATE {\bf Input}: Original dataset $\mathcal{D}$ with features $\boldsymbol{z}_i$. Set of Yara Rules $\mathcal{R}$.\\
\STATE Set of simple signatures $S \gets \emptyset$.
\FOR{Yara Rule $\in \mathcal{R}$}
  \STATE Extract from the Yara Rule all individual sub-signatures $\{r_{j}, \ldots, r_{j'}\}$ 
  \STATE $S \gets S \cup \{r_{j}, \ldots, r_{j'}\} $ {\color{ForestGreen}\COMMENT{ Update the set of signature components}}
\ENDFOR
\FOR{Binary sample  $i \in \mathcal{D}$}
  \STATE $\boldsymbol{\hat{x}}_i \gets $ zero vector of $|S|$ dimensions. 
  \FOR{$r_j \in S$} 
    \STATE $\boldsymbol{\hat{x}}_{i}[j] \gets \mathds{1}\llbracket r_j \in \text{sample}_i\rrbracket$ {\color{ForestGreen} \COMMENT{Set signature occurrence as a one-hot feature}}
  \ENDFOR
  \SWITCH{Adding side information}
    \CASE {Conditional Selection}
      \STATE  $\boldsymbol{\hat{x}}_i \gets [\boldsymbol{\hat{x}}_i ; f_0(\boldsymbol{z}_i) ]$ {\color{ForestGreen} \COMMENT{Add implicit information about original features $\boldsymbol{z}_i$ }}
    \ENDCASE
    \CASE {Stacked feature selection}
      \STATE $\boldsymbol{\hat{x}}_i \gets [\boldsymbol{\hat{x}}_i ; \boldsymbol{z}_i ]$ {\color{ForestGreen} \COMMENT{Explicitly add original features $\boldsymbol{z}_i$ }}
    \ENDCASE
    \DEFAULT
      \STATE $\boldsymbol{\hat{x}}_i$ unaltered {\color{ForestGreen} \COMMENT{Independent selection, no use of side information}}
    \ENDDEFAULT
  \ENDSWITCH
\ENDFOR
\STATE Train selection model $g(\cdot)$ on $\{\boldsymbol{\hat{x}}_1, \boldsymbol{\hat{x}}_2, \ldots, \boldsymbol{\hat{x}}_{|\mathcal{D}|}\}$. 
\STATE $R \gets [r_1', r_2', \ldots, r_k']$ are the top-$k$ relevant Yara features as determined by $g(\cdot)$
\STATE $\boldsymbol{\tilde{x}}_i \gets \boldsymbol{z}_i \cup  \mathds{1}\llbracket r_j \in \text{sample}_i\rrbracket, \forall r_j \in R$, is the final feature set for each sample $i$
\STATE $f(\cdot)$ is the final model trained on $\{\boldsymbol{\tilde{x}}_1, \boldsymbol{\tilde{x}}_2, \ldots, \boldsymbol{\tilde{x}}_{|\mathcal{D}|}\}$
\end{algorithmic}
\caption{\textbf{Ha}rvesting \textbf{Ya}ra rule for \textbf{Ma}lware (HaYaMa)}
\label{alg:hayama}
\end{algorithm}

Our entire approach is summarized in Algorithm \autoref{alg:hayama}, for both the sub-signature extraction and feature selection steps.
Line 10 details the three cases we will evaluate
to decide how side information is added to the model to aid feature selection.
The default case is to ignore that other features $\boldsymbol{z}$ have already been developed and to simply select features based on their predictive performance.
Because this case does not take the prior features into account, it could potentially select redundant features. 

\subsection{Experimental details}

In the following experiments, we use XGBoost models for $f_0$ and $f$. As discussed, they obtained higher accuracy than linear classifiers for a given set of sub-signatures. We use Lasso logistic regression to select features as $g$.

\textbf{Linear models.} We use regularized logistic regression as our feature selection model $g$.
To perform feature selection while fitting the model, we used the $L_1$ regularization (Lasso),
which shrinks some coefficients to zero,
thus identifying the most relevant Yara features for malware classification.
Varying the strength of the regularization parameter $C$ results in varying numbers of selected features.
For this model, we used the LIBLINEAR solver with default parameters~\cite{fan2008liblinear} from scikit-learn~\cite{pedregosa2011scikit}. 
For experiments where the set of features was preset
,
we did not apply the Lasso penalty because we wanted all specified features to be used.

\textbf{Tree-based methods.} We tested XGBoost, LightGBM, and 
Random Forest as our tree-based methods. XGBoost and LightGBM are gradient boosting algorithms, which sequentially adds trees to the model to minimize a conditional objective~\cite{chen2016xgboost,ke2017lightgbm}. 
On the other hand, Random Rorest uses independent randomized trees to minimize decision tree variance~\cite{breiman2001random}.   

We found that XGBoost gave the best results, so we adopted it as our algorithm for $f_0$ and $f$. 
We tuned the hyperparameters for XGBoost using Optuna~\cite{Akiba:2019:ONH:3292500.3330701} with 100 iterations.
The feature set used in our tuning consisted of all EMBER features and 300 YARA rules as identified by Lasso.
The hyperparameter optimization was evaluated using a 20\% validation set from the EMBER training set.
In all experiments, we partition our samples into the original training and test splits provided in the EMBER 2018 dataset \cite{anderson2018ember}.
For additional details and final parameters, refer to the Appendix.

\section{Experimental Results} \label{sec:results}

\subsection{Lasso effectively selects Yara sub-signatures}
We first confirm that Lasso-based selection is more effective than using individual metrics (accuracy, precision, or recall) to filter features.
For any given number of Yara sub-signatures $k$,
we find a Lasso model which keeps approximately $k$ features by choosing $\lambda$ appropriately in Eq.~\ref{eq:lasso}.
For the metric-based approach, we select the top $k$ features as sorted by that metric.

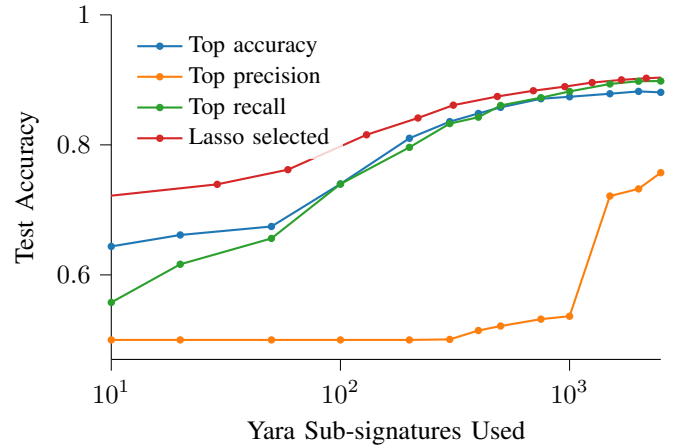
\begin{figure}[!h]
    \centering
    \begin{tikzpicture}

\definecolor{crimson2143940}{RGB}{214,39,40}
\definecolor{darkgray176}{RGB}{176,176,176}
\definecolor{darkorange25512714}{RGB}{255,127,14}
\definecolor{forestgreen4416044}{RGB}{44,160,44}
\definecolor{lightgray204}{RGB}{204,204,204}
\definecolor{steelblue31119180}{RGB}{31,119,180}

\begin{axis}[
width=0.49\textwidth,
height=0.34\textwidth,
axis x line* = bottom,
axis y line* = left,
legend cell align={left},
legend style={
  fill opacity=0.8,
  draw opacity=1,
  text opacity=1,
  at={(0.03,0.97)},
  anchor=north west,
  draw=none
},
log basis x={10},
tick align=outside,
tick pos=left,
x grid style={darkgray176},
xlabel={Yara Sub-signatures Used},
xmin=10, xmax=2500,
xmode=log,
xtick style={color=black},
xtick={0.01,0.1,1,10,100,1000,10000,100000},
xticklabels={
  \(\displaystyle {10^{-2}}\),
  \(\displaystyle {10^{-1}}\),
  \(\displaystyle {10^{0}}\),
  \(\displaystyle {10^{1}}\),
  \(\displaystyle {10^{2}}\),
  \(\displaystyle {10^{3}}\),
  \(\displaystyle {10^{4}}\),
  \(\displaystyle {10^{5}}\)
},
y grid style={darkgray176},
ylabel={Test Accuracy},
ymin=0.47, ymax=1,
ytick style={color=black}
]
\addplot [thick, steelblue31119180, mark=*, mark size=1, mark options={solid}]
table {%
1 0.538235
5 0.632095
10 0.643805
20 0.661405
50 0.67446
100 0.73992
200 0.81017
300 0.83571
400 0.84832
500 0.857705
750 0.870935
1000 0.873985
1500 0.878635
2000 0.88224
2500 0.880775
};
\addlegendentry{\small Top accuracy}
\addplot [thick, darkorange25512714, mark=*, mark size=1, mark options={solid}]
table {%
1 0.5
5 0.5
10 0.5
20 0.5
50 0.5
100 0.5
200 0.5
300 0.50077
400 0.5144
500 0.521365
750 0.531935
1000 0.53644
1500 0.72132
2000 0.732305
2500 0.75724
};
\addlegendentry{\small Top precision}
\addplot [thick, forestgreen4416044, mark=*, mark size=1, mark options={solid}]
table {%
1 0.5
5 0.5
10 0.557645
20 0.61634
50 0.656105
100 0.739705
200 0.796155
300 0.832725
400 0.84256
500 0.860635
750 0.872555
1000 0.88218
1500 0.89368
2000 0.897905
2500 0.898165
};
\addlegendentry{\small Top recall}
\addplot [thick, crimson2143940, mark=*, mark size=1, mark options={solid}]
table {%
1 0.5
1 0.5
7 0.71602
29 0.73931
59 0.761855
130 0.815495
218 0.84132
311 0.86107
484 0.874435
695 0.88326
953 0.8896
1254 0.89583
1684 0.899975
2162 0.902475
2727 0.90404
3436 0.90653
4142 0.906825
4901 0.906765
5512 0.90633
5870 0.90636
};
\addlegendentry{\small Lasso selected}
\end{axis}

\end{tikzpicture}
    \caption{In this experiment, linear classifiers are trained on a subset of Yara sub-signatures to predict Ember test labels. We compare metric-based selection of Yara sub-signatures (filtering by top accuracy, precision, or recall) versus automated feature selection via the Lasso penalty. Lasso selection gives higher accuracy for any feature set size.}
    \label{fig:feature_selection}
\end{figure}

The results are shown in Fig~\ref{fig:feature_selection}.
This experiment has two takeaways: (1) a combination of Yara sub-signatures can achieve significantly higher accuracy than individual features; (2) a linear classifier based on lasso selection obtains higher accuracy for any number of Yara features than metric-based selection. 
This is because the latter approaches do not account for useful interactions among rules and often include rules which individually perform well but are overall redundant.
This is especially important given that the YARA rules do not individually show high discriminative ability.

\subsection{Combining Yara with Prior Features}

We now provide additional detail about our experiments.

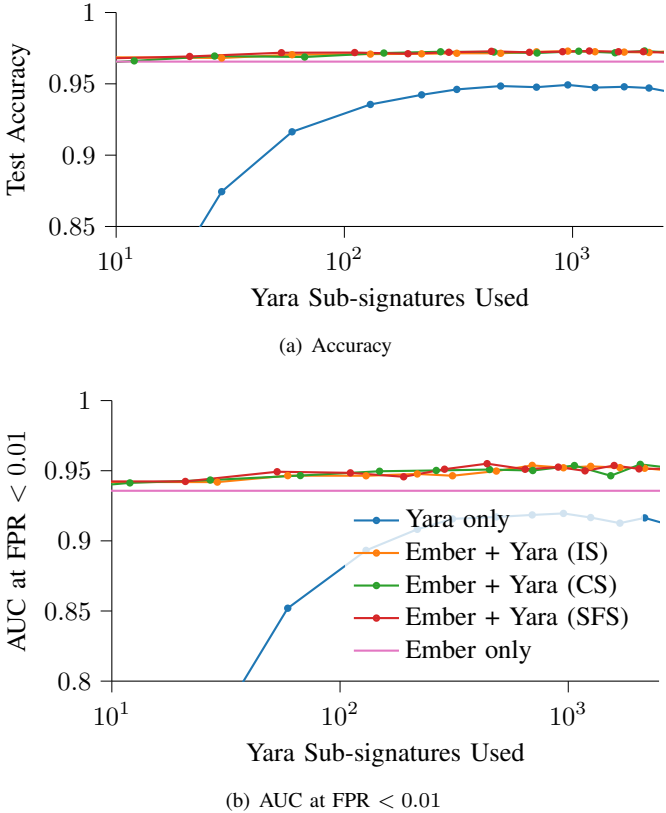
\begin{figure}[!h]
    \centering
    \subfigure[Accuracy]{
    \begin{tikzpicture}

\definecolor{crimson2143940}{RGB}{214,39,40}
\definecolor{darkgray176}{RGB}{176,176,176}
\definecolor{darkorange25512714}{RGB}{255,127,14}
\definecolor{forestgreen4416044}{RGB}{44,160,44}
\definecolor{lightgray204}{RGB}{204,204,204}
\definecolor{orchid227119194}{RGB}{227,119,194}
\definecolor{steelblue31119180}{RGB}{31,119,180}

\begin{axis}[
width=\columnwidth,
height=0.5\columnwidth,
legend cell align={left},
legend style={
  fill opacity=0.8,
  draw opacity=1,
  text opacity=1,
  at={(0.97,0.03)},
  anchor=south east,
  draw=none
},
axis x line* = bottom,
axis y line* = left,
log basis x={10},
tick align=outside,
tick pos=left,
x grid style={darkgray176},
xlabel={Yara Sub-signatures Used},
xmin=10, xmax=2500,
xmode=log,
xtick style={color=black},
xtick={0.1,1,10,100,1000,10000,100000},
xticklabels={
  \(\displaystyle {10^{-1}}\),
  \(\displaystyle {10^{0}}\),
  \(\displaystyle {10^{1}}\),
  \(\displaystyle {10^{2}}\),
  \(\displaystyle {10^{3}}\),
  \(\displaystyle {10^{4}}\),
  \(\displaystyle {10^{5}}\)
},
y grid style={darkgray176},
ylabel={Test Accuracy},
ymin=0.85, ymax=1,
ytick style={color=black}
]
\addplot [thick, steelblue31119180, mark=*, mark size=1, mark options={solid}]
table {%
7 0.71293
29 0.874405
59 0.916365
130 0.93554
218 0.94228
311 0.946085
484 0.948445
695 0.947655
953 0.949245
1254 0.94738
1684 0.94792
2162 0.94707
2727 0.944005
3436 0.94712
4142 0.947605
4901 0.944595
5512 0.947415
5870 0.9469
};
\addplot [thick, darkorange25512714, mark=*, mark size=1, mark options={solid}]
table {%
7 0.96876
29 0.968115
59 0.970405
130 0.970835
218 0.971055
311 0.971425
484 0.971415
695 0.97244
953 0.97293
1254 0.972505
1684 0.972245
2162 0.97198
2727 0.972955
3436 0.97103
4142 0.97131
4901 0.972555
5512 0.9716
5870 0.972135
};
\addplot [thick, forestgreen4416044, mark=*, mark size=1, mark options={solid}]
table {%
2 0.966705
8 0.96663
8 0.96663
9 0.964795
12 0.966105
27 0.969475
67 0.968815
149 0.971585
264 0.97248
452 0.972285
699 0.97161
1064 0.972865
1535 0.971775
2068 0.97306
2760 0.97168
3594 0.972935
4327 0.97323
};
\addplot [thick, crimson2143940, mark=*, mark size=1, mark options={solid}]
table {%
1 0.965565
4 0.966455
21 0.96918
53 0.97181
111 0.97193
190 0.97111
287 0.972125
441 0.972785
647 0.9721
904 0.972555
1184 0.973025
1589 0.972505
2042 0.972545
2574 0.971795
3206 0.97398
3914 0.972565
4688 0.974145
};
\addplot [thick, orchid227119194]
table {%
5 0.96557
5000 0.96557
};
\end{axis}

\end{tikzpicture}
    }
    \subfigure[AUC at FPR $< 0.01$]{
    \begin{tikzpicture}

\definecolor{crimson2143940}{RGB}{214,39,40}
\definecolor{darkgray176}{RGB}{176,176,176}
\definecolor{darkorange25512714}{RGB}{255,127,14}
\definecolor{forestgreen4416044}{RGB}{44,160,44}
\definecolor{lightgray204}{RGB}{204,204,204}
\definecolor{orchid227119194}{RGB}{227,119,194}
\definecolor{steelblue31119180}{RGB}{31,119,180}

\begin{axis}[
width=\columnwidth,
height=0.6\columnwidth,
legend cell align={left},
legend style={
  fill opacity=0.8,
  draw opacity=1,
  text opacity=1,
  at={(0.97,0.03)},
  anchor=south east,
  draw=none
},
axis x line* = bottom,
axis y line* = left,
log basis x={10},
tick align=outside,
tick pos=left,
x grid style={darkgray176},
xlabel={Yara Sub-signatures Used},
xmin=10, xmax=2500,
xmode=log,
xtick style={color=black},
xtick={0.1,1,10,100,1000,10000,100000},
xticklabels={
  \(\displaystyle {10^{-1}}\),
  \(\displaystyle {10^{0}}\),
  \(\displaystyle {10^{1}}\),
  \(\displaystyle {10^{2}}\),
  \(\displaystyle {10^{3}}\),
  \(\displaystyle {10^{4}}\),
  \(\displaystyle {10^{5}}\)
},
y grid style={darkgray176},
ylabel={AUC at FPR $< 0.01$},
ymin=0.8, ymax=1,
ytick style={color=black}
]
\addplot [thick, steelblue31119180, mark=*, mark size=1, mark options={solid}]
table {%
7 0.593422440295
29 0.769264185929648
59 0.85197567839196
130 0.893200386934673
218 0.908230067839196
311 0.915679389447236
484 0.917057610552764
695 0.918480907035176
953 0.919554610552764
1254 0.916615404522613
1684 0.912631118090452
2162 0.916383309045226
2727 0.911741620603015
3436 0.914023082914573
4142 0.915238507537688
4901 0.909919179812104
5512 0.913667633165829
5870 0.91325224120603
};
\addlegendentry{Yara only}
\addplot [thick, darkorange25512714, mark=*, mark size=1, mark options={solid}]
table {%
7 0.941984243718593
29 0.941892331658291
59 0.946496922110553
130 0.946408130653266
218 0.947699680904523
311 0.946427487437186
484 0.94975193718593
695 0.953794650753769
953 0.952106716080402
1254 0.9531251959799
1684 0.952214369346734
2162 0.951762942211055
2727 0.950219874371859
3436 0.950050781407035
4142 0.9452438040201
4901 0.948266060301508
5512 0.946315703517588
5870 0.949411929648241
};
\addlegendentry{Ember + Yara (IS)}
\addplot [thick, forestgreen4416044, mark=*, mark size=1, mark options={solid}]
table {%
2 0.939369298994975
8 0.940358736180904
8 0.940358736180904
9 0.939596173366834
12 0.941324035175879
27 0.94335014321608
67 0.946557613065327
149 0.949623248743719
264 0.950189788944724
452 0.950837600502513
699 0.950135168341709
1064 0.953700535175879
1535 0.946415404522613
2068 0.954529484924623
2760 0.952162311557789
3594 0.954207092964824
4327 0.951557115577889
};
\addlegendentry{Ember + Yara (CS)}
\addplot [thick, crimson2143940, mark=*, mark size=1, mark options={solid}]
table {%
1 0.935699889447236
4 0.942360273869347
21 0.942352610552764
53 0.94927298241206
111 0.948484216080402
190 0.945675806532663
287 0.951076346733668
441 0.955030344221106
647 0.951039806532663
904 0.952567552763819
1184 0.9499093040201
1589 0.953735771356784
2042 0.951309442211055
2574 0.95139590201005
3206 0.951408736180904
3914 0.949842341708543
4688 0.952561037688442
};
\addlegendentry{Ember + Yara (SFS)}
\addplot [thick, orchid227119194]
table {%
5 0.9357
5000 0.9357
};
\addlegendentry{Ember only}
\end{axis}

\end{tikzpicture}
    }
    \caption{Performance of our full model as presented in Algo.~\ref{alg:hayama}, showing how (a) test accuracy, and (b) AUC at low FPR, increases as Yara features are added. A Yara-only model without Ember features (blue) shows the importance of using side information. Furthermore, Yara features demonstrate clear value when added to the baseline Ember-only model (pink).}
    \label{fig:full_model}
\end{figure}

We first demonstrate the performance of our full method as described in Algo.~\ref{alg:hayama} by incorporating the original Ember features. 
As described in section \ref{subsec:side_info}, we modify the feature selection process itself to incorporate information from the Ember features.
These approaches are compared in Fig.~\ref{fig:full_model}, which shows performance in two metrics: test accuracy, and area under ROC (AUC) limited to low false positive rate (FPR) under 0.01.
We find that the approaches give similarly high performance, with the independent selection being marginally lower on occasion.

We also highlight two important observations to demonstrate our method's effectiveness. First, the model using Yara features only (blue) has lower performance than any model which also uses the Ember features, although it still reaches 95\% accuracy. 
This observation supports that side information can provide high predictive value and is important to incorporate into our method.

Second, while the model that only uses original Ember features (pink) achieves a relatively high baseline accuracy of 96.6\%, this performance is immediately surpassed by adding Yara features. Furthermore, the performance continues to increase as more Yara features are added, eventually plateauing at around 300 features. This is similar to the Yara-only curve and supports our claim that \textit{Yara sub-signatures add predictive value beyond what the original Ember features provide}.

\subsection{Effect of Model Choice}

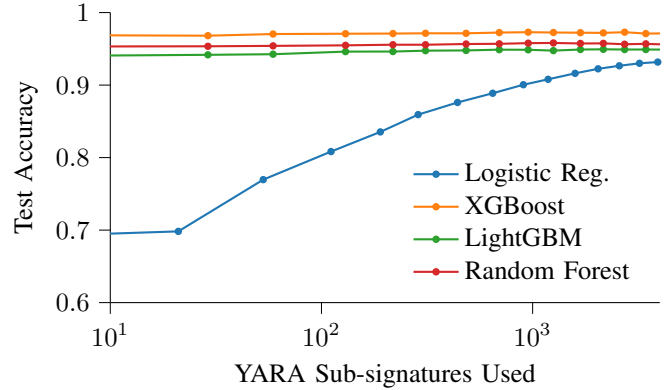
\begin{figure}[!h]
    \centering
    \begin{tikzpicture}

\definecolor{crimson2143940}{RGB}{214,39,40}
\definecolor{darkgray176}{RGB}{176,176,176}
\definecolor{darkorange25512714}{RGB}{255,127,14}
\definecolor{forestgreen4416044}{RGB}{44,160,44}
\definecolor{lightgray204}{RGB}{204,204,204}
\definecolor{steelblue31119180}{RGB}{31,119,180}

\begin{axis}[
width=0.49\textwidth,
height=0.3\textwidth,
legend cell align={left},
legend style={
  fill opacity=0.8,
  draw opacity=1,
  text opacity=1,
  at={(0.97,0.03)},
  anchor=south east,
  draw=none
},
axis x line* = bottom,
axis y line* = left,
log basis x={10},
tick align=outside,
tick pos=left,
x grid style={darkgray176},
xlabel={YARA Sub-signatures Used},
xmin=10, xmax=4000,
xmode=log,
xtick style={color=black},
xtick={1,10,100,1000,10000,100000},
xticklabels={
  \(\displaystyle {10^{0}}\),
  \(\displaystyle {10^{1}}\),
  \(\displaystyle {10^{2}}\),
  \(\displaystyle {10^{3}}\),
  \(\displaystyle {10^{4}}\),
  \(\displaystyle {10^{5}}\)
},
y grid style={darkgray176},
ylabel={Test Accuracy},
ymin=0.6, ymax=1,
ytick style={color=black}
]
\addplot [thick, steelblue31119180, mark=*, mark size=1, mark options={solid}]
table {%
1 0.5
1 0.6081
4 0.69127
21 0.69825
53 0.7696
111 0.808315
190 0.835415
287 0.859245
441 0.876015
647 0.888735
904 0.900495
1184 0.90796
1589 0.91619
2042 0.922445
2574 0.92662
3206 0.930045
3914 0.931745
4688 0.931515
};
\addlegendentry{Logistic Reg.}
\addplot [thick, darkorange25512714, mark=*, mark size=1, mark options={solid}]
table {%
7 0.96876
29 0.968115
59 0.970405
130 0.970835
218 0.971055
311 0.971425
484 0.971415
695 0.97244
953 0.97293
1254 0.972505
1684 0.972245
2162 0.97198
2727 0.972955
3436 0.97103
4142 0.97131
4901 0.972555
5512 0.9716
5870 0.972135
};
\addlegendentry{XGBoost}
\addplot [thick, forestgreen4416044, mark=*, mark size=1, mark options={solid}]
table {%
1 0.93731
7 0.940485
29 0.94175
59 0.942565
130 0.946165
218 0.946265
311 0.947465
484 0.9478
695 0.948845
953 0.94869
1254 0.94755
1684 0.949035
2162 0.94935
2727 0.949035
3436 0.94908
4142 0.94893
4901 0.948825
5512 0.94873
5870 0.9488
};
\addlegendentry{LightGBM}
\addplot [thick, crimson2143940, mark=*, mark size=1, mark options={solid}]
table {%
1 0.95423
7 0.95322
29 0.95352
59 0.954075
130 0.95482
218 0.95567
311 0.955645
484 0.956655
695 0.956925
953 0.95782
1254 0.95811
1684 0.957395
2162 0.957525
2727 0.95634
3436 0.956855
4142 0.95609
4901 0.956235
5512 0.95658
5870 0.9556
};
\addlegendentry{Random Forest}
\end{axis}

\end{tikzpicture}
    \caption{Comparing various machine learning algorithms as YARA rules are \textit{added} to the baseline EMBER features as model training inputs. XGBoost (orange) performs the best across the range of YARA rules used, so we adopt it as our main model.}
    \label{fig:compare_models}
\end{figure}

We also analyze the effect of model choice for our predictive model $f$. In Fig.~\ref{fig:compare_models}, the \textit{independent selection} approach is used to combine Ember features with Yara sub-signatures into various choices of final predictive model. 
The experiment clearly demonstrates the dominance of tree-based methods (XGBoost, LightGBM, and Random Forest) over the linear classification model (logistic regression). In addition, XGBoost has the highest overall performance, which is why we selected it as our final predictive model.

\subsection{Exploration of Selected Features}

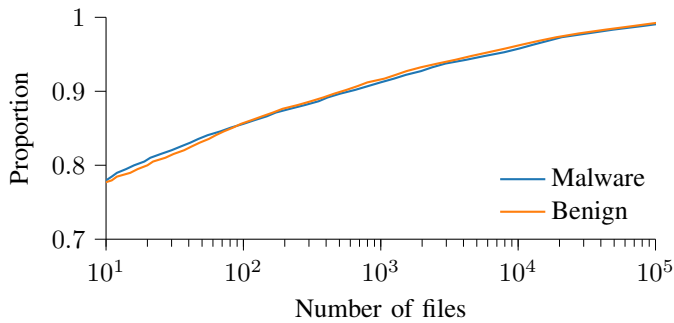
\begin{figure}
    \centering
    \begin{tikzpicture}

\definecolor{darkgray176}{RGB}{176,176,176}
\definecolor{darkorange25512714}{RGB}{255,127,14}
\definecolor{lightgray204}{RGB}{204,204,204}
\definecolor{steelblue31119180}{RGB}{31,119,180}

\begin{axis}[
width=0.49\textwidth,
height=0.25\textwidth,
legend cell align={left},
legend style={
  fill opacity=0.8,
  draw opacity=1,
  text opacity=1,
  at={(0.7,0.37)},
  anchor=north west,
  draw=none
},
axis x line* = bottom,
axis y line* = left,
log basis x={10},
tick align=outside,
tick pos=left,
x grid style={darkgray176},
xmin=10, xmax=100000,
xlabel={Number of files},
xmode=log,
xtick style={color=black},
y grid style={darkgray176},
ymin=0.7, ymax=1,
ylabel={Proportion},
ytick style={color=black}
]
\addplot [thick, steelblue31119180]
table {%
0 0
0 0.00509502216334641
0 0.0101900443266928
0 0.0152850664900392
0 0.0203800886533856
0 0.0254751108167321
0 0.0305701329800785
0 0.0356651551434249
0 0.0407601773067713
0 0.0458551994701177
0 0.0509502216334641
0 0.0560452437968105
0 0.0611402659601569
0 0.0662352881235033
0 0.0713303102868497
0 0.0764253324501962
0 0.0815203546135426
0 0.086615376776889
0 0.0917103989402354
0 0.0968054211035818
0 0.101900443266928
0 0.106995465430275
0 0.112090487593621
0 0.117185509756967
0 0.122280531920314
0 0.12737555408366
0 0.132470576247007
0 0.137565598410353
0 0.142660620573699
0 0.147755642737046
0 0.152850664900392
0 0.157945687063739
0 0.163040709227085
0 0.168135731390432
0 0.173230753553778
0 0.178325775717124
0 0.183420797880471
0 0.188515820043817
0 0.193610842207164
0 0.19870586437051
0 0.203800886533856
0 0.208895908697203
0 0.213990930860549
0 0.219085953023896
0 0.224180975187242
0 0.229275997350588
0 0.234371019513935
0 0.239466041677281
0 0.244561063840628
0 0.249656086003974
0 0.25475110816732
0 0.259846130330667
0 0.264941152494013
0 0.27003617465736
0 0.275131196820706
0 0.280226218984053
0 0.285321241147399
0 0.290416263310745
0 0.295511285474092
0 0.300606307637438
0 0.305701329800785
0 0.310796351964131
0 0.315891374127477
0 0.320986396290824
0 0.32608141845417
0 0.331176440617517
0 0.336271462780863
0 0.34136648494421
0 0.346461507107556
0 0.351556529270902
0 0.356651551434249
0 0.361746573597595
0 0.366841595760942
0 0.371936617924288
0 0.377031640087634
0 0.382126662250981
0 0.387221684414327
0 0.392316706577674
0 0.39741172874102
0 0.402506750904366
0 0.407601773067713
0 0.412696795231059
0 0.417791817394406
0 0.422886839557752
0 0.427981861721098
0 0.433076883884445
0 0.438171906047791
0 0.443266928211138
0 0.448361950374484
0 0.453456972537831
0 0.458551994701177
0 0.463647016864523
0 0.46874203902787
0 0.473837061191216
0 0.478932083354563
0 0.484027105517909
0 0.489122127681255
0 0.494217149844602
0 0.499312172007948
0 0.504407194171295
0 0.509502216334641
0 0.514597238497987
0 0.519692260661334
0 0.52478728282468
0 0.529882304988027
0 0.534977327151373
0 0.540072349314719
0 0.545167371478066
0 0.550262393641412
0 0.555357415804759
0 0.560452437968105
0 0.565547460131452
0 0.570642482294798
0 0.575737504458144
0 0.580832526621491
0 0.585927548784837
0 0.591022570948184
0 0.59611759311153
0 0.601212615274876
0 0.606307637438223
0 0.611402659601569
0 0.616497681764916
1 0.621592703928262
1 0.626687726091608
1 0.631782748254955
1 0.636877770418301
1 0.641972792581648
1 0.647067814744994
1 0.652162836908341
1 0.657257859071687
1 0.662352881235033
1 0.66744790339838
1 0.672542925561726
1 0.677637947725073
1 0.682732969888419
1 0.687827992051765
2 0.692923014215112
2 0.698018036378458
2 0.703113058541805
2 0.708208080705151
2 0.713303102868497
3 0.718398125031844
3 0.72349314719519
3 0.728588169358537
4 0.733683191521883
4 0.738778213685229
4 0.743873235848576
5 0.748968258011922
5 0.754063280175269
6 0.759158302338615
7 0.764253324501961
8 0.769348346665308
9 0.774443368828654
10 0.779538390992001
11 0.784633413155347
12 0.789728435318694
14 0.79482345748204
16 0.799918479645386
19 0.805013501808733
21 0.810108523972079
25 0.815203546135426
30 0.820298568298772
35 0.825393590462118
41 0.830488612625465
47 0.835583634788811
55 0.840678656952158
68 0.845773679115504
81 0.850868701278851
100 0.855963723442197
121 0.861058745605543
148 0.86615376776889
172 0.871248789932236
219 0.876343812095583
281 0.881438834258929
353 0.886533856422275
410 0.891628878585622
505 0.896723900748968
649 0.901818922912315
804 0.906913945075661
995 0.912008967239007
1247 0.917103989402354
1523 0.9221990115657
1965 0.927294033729047
2365 0.932389055892393
2956 0.93748407805574
4184 0.942579100219086
5691 0.947674122382432
7912 0.952769144545779
10227 0.957864166709125
12843 0.962959188872472
16353 0.968054211035818
20781 0.973149233199164
31998 0.978244255362511
49168 0.983339277525857
82089 0.988434299689204
131529 0.99352932185255
252513 0.998624344015896
};
\addlegendentry{Malware}
\addplot [thick, darkorange25512714]
table {%
0 0
0 0.00509502216334641
0 0.0101900443266928
0 0.0152850664900392
0 0.0203800886533856
0 0.0254751108167321
0 0.0305701329800785
0 0.0356651551434249
0 0.0407601773067713
0 0.0458551994701177
0 0.0509502216334641
1 0.0560452437968105
1 0.0611402659601569
1 0.0662352881235033
1 0.0713303102868497
1 0.0764253324501962
1 0.0815203546135426
1 0.086615376776889
1 0.0917103989402354
1 0.0968054211035818
1 0.101900443266928
1 0.106995465430275
1 0.112090487593621
1 0.117185509756967
1 0.122280531920314
1 0.12737555408366
1 0.132470576247007
1 0.137565598410353
1 0.142660620573699
1 0.147755642737046
1 0.152850664900392
1 0.157945687063739
1 0.163040709227085
1 0.168135731390432
1 0.173230753553778
1 0.178325775717124
1 0.183420797880471
1 0.188515820043817
1 0.193610842207164
1 0.19870586437051
1 0.203800886533856
1 0.208895908697203
1 0.213990930860549
1 0.219085953023896
1 0.224180975187242
1 0.229275997350588
1 0.234371019513935
1 0.239466041677281
1 0.244561063840628
1 0.249656086003974
1 0.25475110816732
1 0.259846130330667
1 0.264941152494013
1 0.27003617465736
1 0.275131196820706
1 0.280226218984053
1 0.285321241147399
1 0.290416263310745
1 0.295511285474092
1 0.300606307637438
1 0.305701329800785
1 0.310796351964131
1 0.315891374127477
1 0.320986396290824
1 0.32608141845417
1 0.331176440617517
1 0.336271462780863
1 0.34136648494421
1 0.346461507107556
1 0.351556529270902
1 0.356651551434249
1 0.361746573597595
1 0.366841595760942
1 0.371936617924288
1 0.377031640087634
1 0.382126662250981
1 0.387221684414327
1 0.392316706577674
1 0.39741172874102
1 0.402506750904366
1 0.407601773067713
1 0.412696795231059
1 0.417791817394406
1 0.422886839557752
1 0.427981861721098
1 0.433076883884445
1 0.438171906047791
1 0.443266928211138
1 0.448361950374484
1 0.453456972537831
1 0.458551994701177
1 0.463647016864523
1 0.46874203902787
1 0.473837061191216
1 0.478932083354563
1 0.484027105517909
1 0.489122127681255
1 0.494217149844602
1 0.499312172007948
1 0.504407194171295
1 0.509502216334641
1 0.514597238497987
1 0.519692260661334
1 0.52478728282468
1 0.529882304988027
1 0.534977327151373
1 0.540072349314719
1 0.545167371478066
1 0.550262393641412
1 0.555357415804759
1 0.560452437968105
1 0.565547460131452
1 0.570642482294798
1 0.575737504458144
1 0.580832526621491
1 0.585927548784837
1 0.591022570948184
1 0.59611759311153
1 0.601212615274876
1 0.606307637438223
1 0.611402659601569
1 0.616497681764916
1 0.621592703928262
1 0.626687726091608
1 0.631782748254955
1 0.636877770418301
1 0.641972792581648
1 0.647067814744994
1 0.652162836908341
2 0.657257859071687
2 0.662352881235033
2 0.66744790339838
2 0.672542925561726
2 0.677637947725073
2 0.682732969888419
2 0.687827992051765
2 0.692923014215112
2 0.698018036378458
2 0.703113058541805
2 0.708208080705151
3 0.713303102868497
3 0.718398125031844
3 0.72349314719519
3 0.728588169358537
4 0.733683191521883
4 0.738778213685229
4 0.743873235848576
5 0.748968258011922
6 0.754063280175269
7 0.759158302338615
7 0.764253324501961
8 0.769348346665308
9 0.774443368828654
11 0.779538390992001
12 0.784633413155347
15 0.789728435318694
17 0.79482345748204
20 0.799918479645386
22 0.805013501808733
27 0.810108523972079
31 0.815203546135426
37 0.820298568298772
42 0.825393590462118
48 0.830488612625465
56 0.835583634788811
63 0.840678656952158
72 0.845773679115504
84 0.850868701278851
96 0.855963723442197
115 0.861058745605543
137 0.86615376776889
164 0.871248789932236
193 0.876343812095583
249 0.881438834258929
310 0.886533856422275
381 0.891628878585622
457 0.896723900748968
561 0.901818922912315
676 0.906913945075661
799 0.912008967239007
1063 0.917103989402354
1284 0.9221990115657
1542 0.927294033729047
1953 0.932389055892393
2590 0.93748407805574
3494 0.942579100219086
4553 0.947674122382432
6081 0.952769144545779
8097 0.957864166709125
10595 0.962959188872472
14040 0.968054211035818
19072 0.973149233199164
27711 0.978244255362511
43301 0.983339277525857
70096 0.988434299689204
111878 0.99352932185255
232999 0.998624344015896
};
\addlegendentry{Benign}
\end{axis}

\end{tikzpicture}
    \caption{The Empirical Cumulative Distribution Function (ECDF) of selected sub-signatures in our training data shows what proportion of the sub-signatures (y-axis) have occurred in how many files (x-axis). We can see that many features selected are highly specific (occur in $\leq 10$ files, left-side), while there are also features that occur in nearly all 600k training files. This power law-esque distribution of features is similar to observations in n-gram-based feature models, showing that our sub-signatures are not too dissimilar in their occurrence rate as previously considered features. }
    \label{fig:Yara_Occurance_Counts}
\end{figure}

Having demonstrated empirically that the sub-signatures found by our approach meaningfully improve the classification accuracy of Ember at both normal and low false-positive scenarios, we further explore the nature of individual-specific features that our model tends to select. First, is it is worth noting that the features selected tend to follow a power-law distribution in their occurrence rate, similar to that of n-gram features~\cite{raff_investigation_2018}. This is shown by the Empirical Cumulative Distribution Function (ECDF) of the sub-signature occurrence rate in both benign and malicious files in \autoref{fig:Yara_Occurance_Counts}. 

Below, we will briefly review some of the features found in our approach, grouped into three common types of selected features. There are two common trends we observe across all cases:
\begin{enumerate}
\item
Features that are highly specific in what they identify (e.g., a specific malware family). It is likely that these features serve a similar purpose as the highly specific ``Kilo-Grams'' of~\cite{Raff2019c,hashgram_2018,raff_hash_gram_parallel}, where 64 to 1024 byte identifiers provided utility when added to the original EMBER features. The hypothesis purposed in ~\cite{Raff2019c} was that such specific identifiers allowed a tree-based model (like XGBoost) to quickly isolate/detect ``problem'' families and then reserve feature splits/utility to more accurately separate the larger populations. We believe this insight applies to our experiments as well. 

\item
Features that are general purpose in nature. These features capture specific behaviors/actions that may occur in a wider class of malware samples or even serve a dual-use purpose that causes them to appear in benign files as well. 
\end{enumerate}

Three common groupings we have identified are file paths, ``atomic strings'', and DLLs. Some examples with an exposition on interesting cases are provided below. Generally, the path features tended to be more specific in nature, atomic strings more diverse in their potential utility, and DLLs almost always are potentially dual-use. The following sub-sections will provide a brief summary of some exemplar sub-signatures we found interesting, as well as a list of strings selected by our approach for inspection.

\subsubsection{Paths}

A common issue in AV deployments is the selection of relevant signatures.   %
With a limited amount of space and compute resources available, it is often the case that signatures are used to tackle the newest threats demanded by customers. The \mintinline[]{text}{ativpsrz.bin} is an example of our approach automatically performing this work by identifying the SPHINX MOTH malware from 2015. Yet all malware data in EMBER 2018 has a VirusTotal first seen date of at least 2018. This is useful from a practical perspective of automating existing procedures for detections while also being potentially lighter weight than keeping/updating a complete version of the original SPING MOTH signature. That is to say, a sub-signature in conjunction with existing features $\boldsymbol{z}$ may be sufficient and have fewer total components than the original signature. Similarly, \mintinline[]{text}{oem7A.PNF} identifies some Stuxnet malware~\cite{Langner2011} and the ``Ben'' path is a case of a malware author seemingly leaving their own identifying information in a build of the ChurnyRoll malware\footnote{\url{https://bartblaze.blogspot.com/2017/11/crunchyroll-hack-delivers-malware.html}}.

Other features like the path to \texttt{Wordpade.exe} are often abused by malware\footnote{\url{https://www.bleepingcomputer.com/news/security/qbot-malware-abuses-windows-wordpad-exe-to-infect-devices/}} 
as an exploit, obfuscation, or attack vector. In this instance, our approach has harvested a generically useful malware detector that is not individually specific.

\begin{minted}[linenos=false,breaklines,breakanywhere,fontsize=\small]{text}
"Lcom/metasploit/stage/c;"
"%
"%
"\\Registry\\Machine\\SYSTEM\\CurrentControlSet\\Control\\Class\\{4D36E972-E325-11CE-BFC1-08002BE10318"
"Users\\Wool3n.H4t\\"
"C:\\Users\\Ben\\Desktop\\taiga-develop\\bin\\Debug\\Taiga.pdb"
"%
\end{minted}

\subsubsection{Atomic Strings}

The ``atomic'' strings in the signatures have a wide variety in content and apparent general utility. For example the \mintinline[]{text}{blazagne.exe.manifest} string indicates use of open source exploitation software for password finding \footnote{\url{https://github.com/AlessandroZ/LaZagne}}, providing us an indicator that this specific tool is in wide enough use that our model found it worthwhile to select as a predictive feature. In this way we have automatically performed the kind of manual engineering work that might have been done to improve a model, without having to do the upfront research to determine that this software was in wide use. 
Similarly
\texttt{return fsdjkl;} is used by the Eleonore Exploit Kit, helping to cover a range of malicious files. 

There are also many dual-use cases of strings that can have both benign and malicious uses. \mintinline[breaklines,breakanywhere]{text}{SELECT Description FROM Win32\_VideoController} is a potential example of a dual-purpose feature that can be used for both benign and malicious intent. The string is a common identifier of an operating system running within a Virtual Machine (VM). This may be used for compatibility reasons in benign software or by malware that is attempting to subvert detection.  \mintinline[]{text}{_crt_debugger_hook} is a general code API often used by both malware and benign applications. 

\begin{minted}[linenos=false,breaklines,breakanywhere,fontsize=\small]{text}
"SELECT Description FROM Win32_VideoController"
"!@#%
"Ljpltmivvdcbb"
"szFileUrl=%
"!! Use Splice Socket !!"
"%
"# \\u@\\h:\\w \\$"
"_crt_debugger_hook"
"[.] Sending shellcode to inject DLL"
"Game Over Good Luck By Wind"
"zhoupin exploit crew"
"\\\\.\\%s"
"L$,PQR"
"1002=/c del /q %
"L|-1|try to run dll %
"%
">>>AUTOIT NO CMDEXECUTE<<<"
"\"%
"return fsdjkl;"
"itifdf)d1L2f'asau%
"blazagne.exe.manifest"
"DontrolService"
"Run the command normally (without code injection)"
"http://www.pretentiousname.com/misc/win7_uac_whitelist2.html"
"*** LOG INJECTS *** %
"lld.trcvsm"
"sys_tkill"
"!deactivebc"
"nView_skins"
"OSQLPASSWORD"
"\\ziedpirate.ziedpirate-PC\\"
"need to do is submit the payment and get the decryption password."
"killOpenedProcessTree"
"cmd.exe /cREG add HKCU\\Software\\Microsoft\\Windows\\CurrentVersion\\Policies\\ActiveDesktop /v NoChangingWallPaper /t REG_DWOR"
"We Control Your Digital World"
"Proactive Bot Killer Enabled!"
"PredatorLogger"
\end{minted}

\subsubsection{dll}
The final list of dynamically linked libraries (DLL) is a straightforward demonstration of various libraries in use for both benign and malicious tasks.  

\begin{minted}[linenos=false,breaklines,breakanywhere,fontsize=\small]{text}
"msgrthlp.dll", "gdiplus.dll", "NSCortr.dll"
"\x00Screen.dll", "GDI32.DLL", "DelphiNative.dll"
\end{minted}

\subsection{Confirming the Additional Information of Yara sub-signatures}

\begin{figure}[!h]
    \centering
    \adjustbox{max width=0.8\columnwidth}{%
        \includegraphics[]{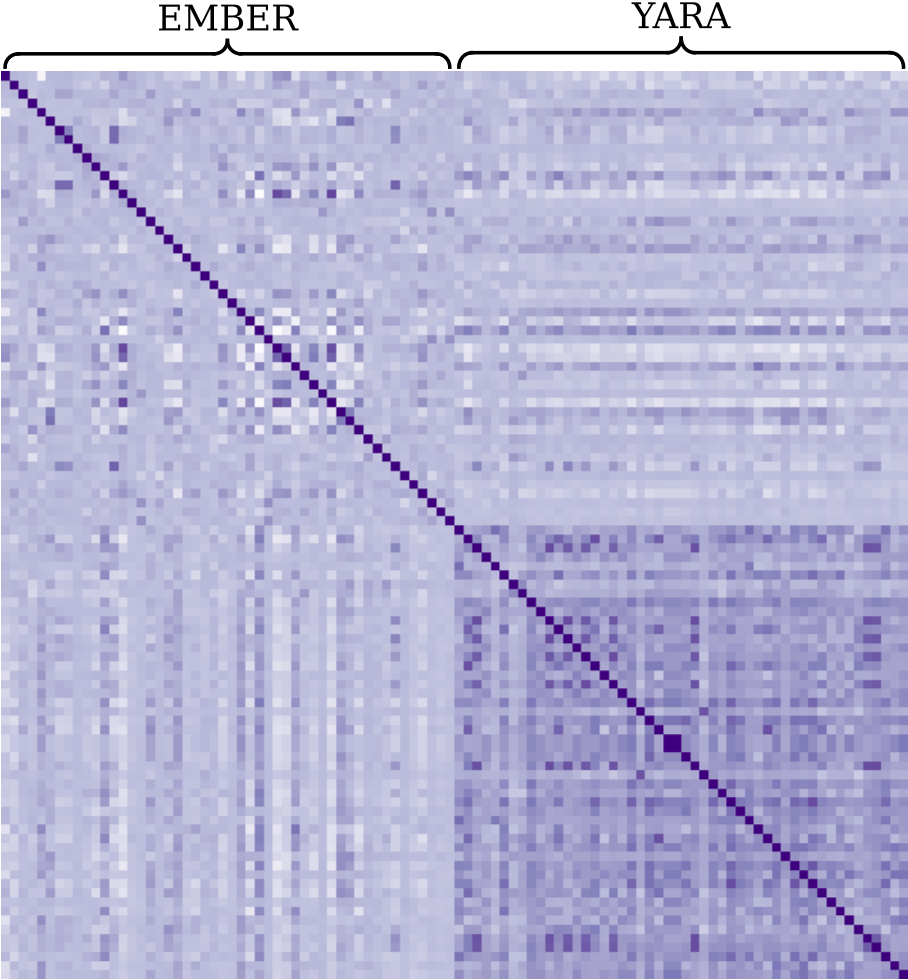}
    }
    \caption{Inter-correlation matrix among top 50 Ember and Yara features as determined by feature selection. Darker values are higher correlation. This shows that the Yara sub-signatures have no special correlation with the original EMBER features, and thus, we are selecting useful new information. }
    \label{fig:corrmap}
\end{figure}

Because the EMBER features already contain string and header information, it is worth further confirming that the sub-signatures selected are meaningfully distinct from the already existing EMBER features $\boldsymbol{z}$. A first pass at this can be done by looking at the inter-correlation between the top features from the original EMBER and the top YARA sub-signatures, as shown in \autoref{fig:corrmap}. 

The original EMBER includes seven general categories of features, and so it is reasonable that the Ember portion of the intercorrelation matrix shows sub-regions of correlation. By contrast, we see that the Yara sub-signatures have much higher correlation values. The regular correlation of these features is a good indicator of why non-linear models are necessary to obtain good performances from the Yara sub-signatures (i.e., co-linear features are intrinsically redundant in a linear model). The important result is that the Yara and EMBER portions have a relatively low correlation. 

\begin{figure}[!h]
    \centering
    \begin{tikzpicture}

\definecolor{darkgray176}{RGB}{176,176,176}
\definecolor{steelblue31119180}{RGB}{31,119,180}

\begin{axis}[
width=0.49\textwidth,
height=0.25\textwidth,
log basis x={10},
tick align=outside,
tick pos=left,
x grid style={darkgray176},
axis x line* = bottom,
axis y line* = left,
xlabel={Yara Sub-signatures Used},
xmin=50, xmax=1000,
xmode=log,
xtick style={color=black},
y grid style={darkgray176},
ylabel={Max correlation},
ymin=0.550263030669925, ymax=0.732088436435424,
ytick style={color=black}
]
\addplot [very thick, steelblue31119180, mark=*, mark size=1, mark options={solid}]
table {%
7 0.558527821841084
29 0.599297994606739
59 0.648006800970042
130 0.692184526580395
218 0.698491521983663
311 0.704292178500151
484 0.706892224863924
695 0.712806614750317
953 0.718561134128255
1254 0.718224342936975
1684 0.721182459793152
2162 0.723776915030193
2727 0.723823645264265
};
\end{axis}

\end{tikzpicture}
    \caption{Maximal correlation between selected Yara sub-signatures with the set of 100 most predictive Ember features, as determined through partial least squares regression. This shows that our approach provides new information independent of the original EMBER features.}
    \label{fig:pls}
\end{figure}
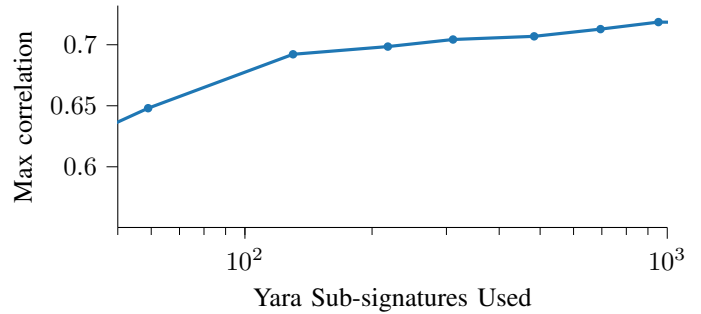

A more precise quantification of this is the maximal correlation coefficient, which we calculate as a function of the number of sub-signatures used in \autoref{fig:pls}. The correlation is non-trivial and plateaus at 70\%. That is to say, 70\% of the original EMBER feature information could be linearly reconstructed from the Yara sub-signatures, and the other 30\% is (effectively) new information. 

This is again a positive indicator of our approach to finding new information. Some level of correlation is expected due to the EMBER feature extractor already considering the same types of information (paths, strings, imports) that are prevalent in the selected sub-signatures. This provides a more precise quantification of how much more information is being derived, providing the non-trivial lift in predictive accuracy.

\section{Conclusion} \label{sec:conclusion}

Malware detection requires continuous updating of its features and methods to handle the adversarial drift that occurs in practice. Yet, malware analysts are overworked and busy. Often, their job requires the creation of signatures to provide additional protection against current threats not being captured by the general-purpose component of anti-virus systems. Our work has shown that it is possible to harmonize these conflicting work streams. Yara rules can be collected in bulk from analyst work, and meaningful features can be extracted and incorporated into a production model. This yields a 1.8\% relative boost in performance at no additional work for the analysts. 

\printbibliography

\appendix

\section{Yara Repositories} \label{sec:yara_repositories}

The following is a list of GitHub repositories which we found with publicly available YARA rules. We downloaded the available YARA rules in order to extract sub-signatures.

\begin{itemize}
    \item \url{https://github.com/reversinglabs/reversinglabs-YARA-rules} 
    \item \url{https://github.com/InQuest/awesome-YARA}
    \item \url{https://github.com/AlienVault-Labs/AlienVaultLabs/tree/master/malware\_analysis}
    \item \url{https://gist.github.com/pedramamini/c586a151a978f971b70412ca4485c491}
    \item \url{https://github.com/bwall/bamfdetect/blob/master/BAMF\_Detect/modules/YARA/xtreme.YARA}
    \item \url{https://github.com/bartblaze/YARA-rules}
    \item \url{https://github.com/airbnb/binaryalert/tree/master/tests}
    \item \url{https://github.com/codewatchorg/Burp-YARA-Rules}
    \item \url{https://github.com/kevoreilly/CAPEv2/}
    \item \url{https://github.com/CyberDefenses/CDI\_YARA}
    \item \url{https://github.com/citizenlab/malware-signatures}
    \item \url{https://github.com/MalGamy/YARA\_Rules}
    \item \url{https://github.com/eset/malware-ioc}
    \item \url{https://github.com/kevoreilly/CAPE}
    \item \url{https://github.com/citizenlab}
    \item \url{https://github.com/stvemillertime/ConventionEngine}
    \item \url{https://github.com/deadbits/YARA-rules}
    \item \url{https://github.com/DidierStevens/DidierStevensSuite}
    \item \url{https://github.com/elastic/protections-artifacts/tree/main}
    \item \url{https://github.com/fideliscyber/indicators/tree/master/YARArules}
    \item \url{https://github.com/mandiant/red\_team\_tool\_countermeasures/tree/master}
    \item \url{https://github.com/Neo23x0/signature-base/tree/master}
\end{itemize}

\end{document}